\documentclass{article}
\usepackage{spconf,amsmath,graphicx,hyperref}
\usepackage{fontawesome5}

\title{Joint Optimization of ASV and CM tasks: BTUEF Team's Submission for WildSpoof Challenge}
\name{O\u{g}uzhan Kurnaz$^{1}$, Jagabandhu Mishra$^{2}$, Tomi Kinnunen$^{2}$, Cemal Hanil\c{c}i$^{1}$}
\address{$^{1}$ Electrical and Electronics Engineering, Bursa Technical University \\ $^{2}$ School of Computing, University of Eastern Finland}

\begin{document}
%
\maketitle
\begin{abstract}

Spoofing-aware speaker verification (SASV) jointly addresses automatic speaker verification and spoofing countermeasures to improve robustness against adversarial attacks. In this paper, we investigate our recently proposed modular SASV framework that enables effective reuse of publicly available ASV and CM systems through non-linear fusion, explicitly modeling their interaction, and optimization with an operating-condition-dependent trainable a-DCF loss. The framework is evaluated using ECAPA-TDNN and ReDimNet as ASV embedding extractors and SSL-AASIST as the CM model, with experiments conducted both with and without fine-tuning on the WildSpoof SASV training data. Results show that the best performance is achieved by combining ReDimNet-based ASV embeddings with fine-tuned SSL-AASIST representations, yielding an a-DCF of $0.0515$ on the progress evaluation set and $0.2163$ on the final evaluation set.

\end{abstract}

\section{Introduction}

Automatic speaker verification (ASV) \cite{reynolds94_asriv} systems are widely used in security-critical applications, yet remain vulnerable to spoofing attacks such as replay \cite{muller25_interspeech}, text-to-speech, and voice conversion \cite{das20c_interspeech}. Although dedicated countermeasure (CM) systems can detect spoofed speech, they do not verify speaker identity, making standalone ASV or CM systems insufficient under adversarial conditions \cite{kurnaz24_asvspoof, buker25_interspeech}. This has motivated spoofing-aware speaker verification (SASV), which jointly addresses speaker verification and spoof detection. Recent evaluation campaigns, including SASV2022 challenge \cite{jung2022sasv}, ASVspoof~5 (Track~2) challenge \cite{Wang2024_ASVspoof5}, and the WildSpoof challenge \cite{wu2025wildspoof}, have further emphasized the need for flexible and robust SASV architectures capable of generalizing to unconstrained spoofing scenarios.


In this work, we extend our recently proposed unified SASV framework~\cite{kurnaz2025joint} and use it as a controlled experimental testbed to systematically analyze embedding-level design choices in spoofing-aware speaker verification. Specifically, we quantitatively evaluate the impact of different ASV and CM embedding extractors under both frozen and fine-tuned training regimes. All configurations operate within a calibrated non-linear score-level fusion framework ~\cite{wang24l_interspeech} optimized using a joint SASV objective defined by user-specified operating conditions ~\cite{kurnaz2024optimizing}.

\vspace{-0.2cm}
\section{Proposed Method}

\begin{figure}[t]
  \centering
  \hspace*{-0.05\linewidth}
  \includegraphics[
    width=1.15\linewidth,
    trim=0pt 190pt 0pt 100pt,
    clip
  ]{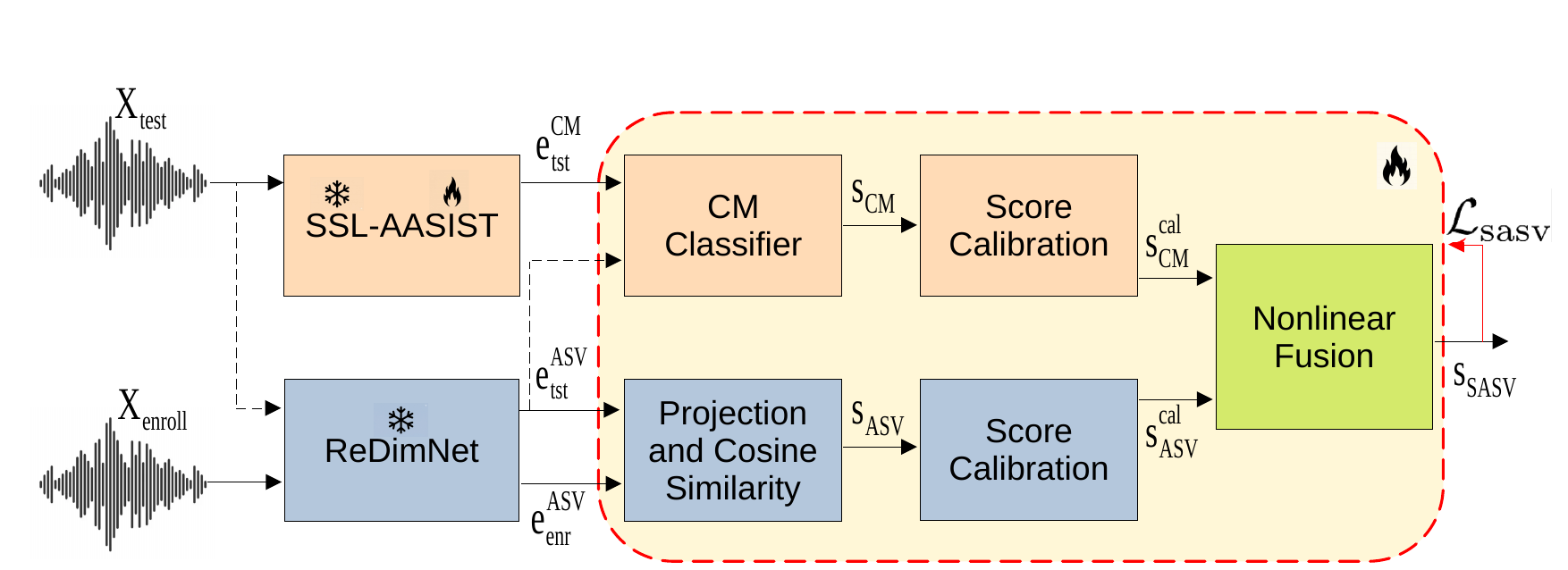}
  \caption{%
  Proposed modular SASV architecture with trainable score-level integration. The pretrained ReDimNet-based ASV extractor is frozen, while only the AASIST component is trained in the SSL-AASIST branch with the SSL encoder frozen. ASV and CM embeddings are converted into calibrated scores, and all modules inside the red dashed block are trainable, producing the final SASV score via nonlinear fusion.}
  \label{fig:proposed_approach}
  \vspace{-0.3cm}
\end{figure}

As illustrated in Fig.~\ref{fig:proposed_approach}, the proposed system consists of three main components—an
ASV branch, a CM branch, and a score fusion module.

\textbf{ASV branch:} Given enrollment and test utterances, $\mathbf{X}_{\mathrm{enr}}$ and
$\mathbf{X}_{\mathrm{tst}}$, fixed ASV encoders extract speaker embeddings
$\mathbf{e}_{\mathrm{enr}}^{\mathrm{asv}}$ and
$\mathbf{e}_{\mathrm{tst}}^{\mathrm{asv}}$. Speaker similarity is computed using
a weighted cosine similarity, where a learnable vector
$\mathbf{w}_{\mathrm{asv}}$ reweights embedding dimensions as
$\mathbf{e}_1=\mathbf{w}_{\mathrm{asv}}\odot\mathbf{e}_{\mathrm{enr}}^{\mathrm{asv}}$
and
$\mathbf{e}_2=\mathbf{w}_{\mathrm{asv}}\odot\mathbf{e}_{\mathrm{tst}}^{\mathrm{asv}}$.
The resulting ASV score is affine-calibrated as
$s_{\mathrm{asv}}^{\mathrm{cal}} = w_0^{\mathrm{asv}} + w_1^{\mathrm{asv}}
s_{\mathrm{asv}}$, yielding an interpretable log-likelihood ratio (LLR) suitable
for fusion.

\textbf{CM branch:} To incorporate spoofing awareness, a CM embedding
$\mathbf{e}_{\mathrm{tst}}^{\mathrm{cm}}$ is extracted from the test utterance
using a frozen CM encoder. This embedding is concatenated with the ASV test
embedding to form a joint representation
$\mathbf{e}_{\mathrm{fused}}=
[\mathbf{e}_{\mathrm{tst}}^{\mathrm{asv}};\mathbf{e}_{\mathrm{tst}}^{\mathrm{cm}}]$,
which is passed through an MLP-based CM classifier. The resulting CM score is
affine-calibrated as
$s_{\mathrm{cm}}^{\mathrm{cal}} = w_0^{\mathrm{cm}} + w_1^{\mathrm{cm}}
s_{\mathrm{cm}}$, ensuring compatibility with the ASV LLR.

\textbf{Score fusion:} The calibrated ASV and CM LLRs are combined via a non-linear score-level fusion
that explicitly models their interaction,
\begin{equation}
s_{\mathrm{sasv}} = -\log\!\left[
(1-\tilde{\rho})e^{-s_{\text{asv}}^{\text{cal}}}
+ \tilde{\rho}e^{-s_{\text{cm}}^{\text{cal}}}
\right],
\label{eq:nonlinearfusion}
\end{equation}
where $\tilde{\rho}\in[0,1]$ controls the relative contribution of speaker and
spoofing evidence. This formulation produces a single SASV score reflecting
joint confidence in target identity and bonafide speech.

\textbf{Joint optimization:} All trainable components—embedding reweighting, calibration layers, CM
classifier, and fusion module—are optimized end-to-end at the SASV decision
level. Each trial is labeled positive only if it is both target and bonafide.
Training uses a weighted combination of binary cross-entropy and the
architecture-agnostic detection cost function (a-DCF) \cite{shim2024adcf}, enabling direct
alignment with SASV evaluation criteria while maintaining a modular system
design.

\vspace{-0.5cm}

\section{Experimental Setup}

In the scope of the competition, the baseline models officially provided by the organizers—namely \textbf{ECAPA-TDNN}, \textbf{MFA-Conformer}, and \textbf{SKA-TDNN}\footnote{Baseline models GitHub repository: \url{https://github.com/wildspoof/SASV_baselines}}—are first trained on the WildSpoof dataset to obtain reference results. 

Building upon these baseline results, state-of-the-art ASV models are integrated into the proposed unified SASV framework. For speaker embedding extraction, we evaluate the \textbf{ECAPA-TDNN}\footnote{ECAPA-TDNN GitHub repository: \url{https://github.com/TaoRuijie/ECAPA-TDNN}} and \textbf{ReDimNet}\footnote{ReDimNet GitHub repository: \url{https://github.com/IDRnD/redimnet}} architectures under three ASV configurations: (i) a \emph{pretrained} and frozen ECAPA-TDNN, (ii) an ECAPA-TDNN model \emph{fine-tuned on the WildSpoof dataset}, and (iii) a \emph{pretrained} and frozen ReDimNet. Speaker embeddings are extracted from the WildSpoof dataset using each configuration, enabling a controlled comparative evaluation within a unified SASV formulation.

For the CM branch, we employ the \textbf{SSL-AASIST}\footnote{SSL-AASIST GitHub repository: \url{https://github.com/TakHemlata/SSL_Anti-spoofing}} model. In this setup, only the AASIST-based spoofing detection component is fine-tuned using the training partition of the WildSpoof dataset, while the self-supervised front-end remains frozen. The resulting CM embeddings are then used within the proposed SASV architecture.

The proposed SASV back-end is implemented as a modular network consisting of two fully connected hidden layers with $384$ and $160$ neurons, respectively. The entire SASV model is trained for $300$ epochs using a batch size of $192$ and a learning rate of $0.005$. The value of $\tilde{\rho}$ in Eq.~\ref{eq:nonlinearfusion} is set to $0.5$. These hyperparameters are kept fixed across all experiments to ensure fair comparison between different ASV embedding backbones and fusion configurations.

\vspace{-0.2cm}
\section{Results}

The ECAPA-TDNN, multi-scale feature aggregation conformer (MFA-Conformer), and selective kernel attention based TDNN (SKA-TDNN) models correspond to the official baseline systems provided by the organizers and are trained strictly following their original configurations, yielding the reference development-set a-DCF results of $0.5113$, $0.4453$, and $0.3118$, respectively, as reported in
Table~\ref{tab:combined_results}. Among these baselines, SKA-TDNN achieves the best development-set performance, and its evaluation-set a-DCF is subsequently computed as $0.3821$. The organizer-provided ECAPA-TDNN is then trained and combined with a pretrained SSL-AASIST model trained on ASVspoof~5, resulting in a development a-DCF of $0.1193$, which is further reduced to $0.0985$ when the
SSL-AASIST component is fine-tuned. Additional experiments using fine-tuned SSL-AASIST embeddings with either a pretrained ECAPA-TDNN or a pretrained ReDimNet backbone yield development a-DCF values of $0.1500$ and $0.0470$, respectively. The best-performing system, based on pretrained ReDimNet and fine-tuned SSL-AASIST, achieves an a-DCF of $0.0515$ on the evaluation set and
$0.2163$ on the final evaluation set.

\vspace{-0.3cm}

\begin{table}[h]
\footnotesize
\centering
\caption{Combined development, evaluation, and final evaluation a-DCF results.
Baseline systems are shown in the upper block, while proposed SASV configurations are listed in the lower block.
T: trained (from scratch), PT: pretrained, FT: finetuned.}
\label{tab:combined_results}
\begin{tabular}{l c c c}
\hline
\textbf{Model} & \textbf{Dev} & \textbf{Eval} & \textbf{Final Eval} \\
\hline
\multicolumn{4}{l}{\textit{Baseline systems}} \\
\hline
ECAPA-TDNN & 0.5113 & N/A & N/A \\
MFA-Conformer & 0.4453 & N/A & N/A \\
SKA-TDNN & 0.3118 & 0.3821 & N/A \\
\hline
\multicolumn{4}{l}{\textit{Proposed SASV}} \\
\hline
T-ECAPA + PT-SSL-AASIST & 0.1193 & N/A & N/A \\
T-ECAPA + FT-SSL-AASIST & 0.0985 & N/A & N/A \\
PT-ECAPA + FT-SSL-AASIST & 0.1500 & N/A & N/A \\
PT-ReDimNet + FT-SSL-AASIST & \textbf{0.0470} & \textbf{0.0515} & \textbf{0.2163} \\
\hline
\end{tabular}
\end{table}

\vspace{-0.6cm}

\section{Conclusion}

We investigated the baseline systems and observed that SKA-TDNN performs best among them. The proposed system with pretrained ECAPA-TDNN and SSL-AASIST outperforms the baselines, with further improvements achieved through fine-tuning. Additional gains are obtained by replacing ECAPA-TDNN with ReDimNet as the ASV embedding extractor. Performance on the final evaluation set is notably lower than on the eval set, possibly due to increased difficulty or unseen conditions, and will be further analyzed once the final evaluation labels are released.

\vfill\pagebreak

\bibliographystyle{IEEEbib}
\bibliography{bibliography}

@article{wu2025wildspoof,
  title={Wildspoof challenge evaluation plan},
  author={Wu, Yihan and Jung, Jee-weon and Shim, Hye-jin and Cheng, Xin and Wang, Xin},
  journal={arXiv preprint arXiv:2508.16858},
  year={2025}
}

@inproceedings{muller25_interspeech,
  title     = {{Replay Attacks Against Audio Deepfake Detection}},
  author    = {Nicolas Müller and Piotr Kawa and Wei-Herng Choong and Adriana Stan and Aditya Tirumala Bukkapatnam and Karla Pizzi and Alexander Wagner and Philip Sperl},
  booktitle = {Proc. {Interspeech 2025}},
  pages     = {2245--2249},
  doi       = {10.21437/Interspeech.2025-20},
  issn      = {2958-1796},
}

@inproceedings{das20c_interspeech,
  title     = {The Attacker’s Perspective on Automatic Speaker Verification: An Overview},
  author    = {Rohan Kumar Das and Xiaohai Tian and Tomi Kinnunen and Haizhou Li},
  booktitle = {Proc. Interspeech 2020},
  pages     = {4213--4217},
  doi       = {10.21437/Interspeech.2020-1052},
  issn      = {2958-1796},
}

@inproceedings{buker25_interspeech,
  title     = {{Evaluating Parameter Sharing for Spoofing-Aware Speaker Verification: A Case Study on the ASVspoof 5 Dataset}},
  author    = {Aykut Büker and Oğuzhan Kurnaz and Şule Bekiryazıcı and Selim Can Demirtaş and Cemal Hanilçi},
  booktitle = {{Proc. Interspeech 2025}},
  pages     = {4573--4577},
  doi       = {10.21437/Interspeech.2025-2618},
  issn      = {2958-1796},
}

@inproceedings{shim2024adcf,
  title     = {{a-DCF}: an architecture agnostic metric with application to spoofing-robust speaker verification},
  author    = {Hye-jin Shim and Jee-weon Jung and Tomi Kinnunen and Nicholas Evans and Jean-François Bonastre and Itshak Lapidot},
  booktitle = {Proc. Odyssey 2024},
  pages     = {158--164},
  doi       = {10.21437/odyssey.2024-23},
}

@inproceedings{wang24l_interspeech,
  title     = {Revisiting and Improving Scoring Fusion for Spoofing-aware Speaker Verification Using Compositional Data Analysis},
  author    = {Xin Wang and Tomi Kinnunen and Kong Aik Lee and Paul-Gauthier Noé and Junichi Yamagishi},
  booktitle = {Proc. Interspeech 2024},
  pages     = {1110--1114},
  doi       = {10.21437/Interspeech.2024-422},
  issn      = {2958-1796},
}

@inproceedings{reynolds94_asriv,
  author={Douglas A. Reynolds},
  title={{Speaker identification and verification using Gaussian mixture speaker models}},
  year=1994,
  booktitle={Proc. ESCA Workshop on Automatic Speaker Recognition, Identification and Verification},
  pages={27--30}
}

@ARTICLE{kurnaz2024optimizing,
  author={Kurnaz, Oğuzhan and Mishra, Jagabandhu and Kinnunen, Tomi H. and Hanilçi, Cemal},
  journal={IEEE Signal Processing Letters}, 
  title={Optimizing a-DCF for Spoofing-Robust Speaker Verification}, 
  year={2025},
  volume={32},
  number={},
  pages={1081-1085},
  doi={10.1109/LSP.2025.3545290}}

@inproceedings{Wang2024_ASVspoof5,
  title     = {{ASV}spoof 5: crowdsourced speech data, deepfakes, and adversarial attacks at scale},
  author    = {Xin Wang and Héctor Delgado and Hemlata Tak and Jee-weon Jung and Hye-jin Shim and Massimiliano Todisco and Ivan Kukanov and Xuechen Liu and Md Sahidullah and Tomi H. Kinnunen and Nicholas Evans and Kong Aik Lee and Junichi Yamagishi},
  booktitle = {Proc. The Automatic Speaker Verification Spoofing Countermeasures Workshop (ASVspoof 2024)},
  pages     = {1--8},
  doi       = {10.21437/ASVspoof.2024-1},
}

@inproceedings{jung2022sasv,
  author={Jee-weon Jung and Hemlata Tak and Hye-jin Shim and Hee-Soo Heo and Bong-Jin Lee and Soo-Whan Chung and Ha-Jin Yu and Nicholas Evans and Tomi Kinnunen},
  title={{SASV 2022: The First Spoofing-Aware Speaker Verification Challenge}},
  booktitle={Proc. Interspeech 2022},
  pages={2893--2897},
  doi={10.21437/Interspeech.2022-11270},
  issn={2958-1796}
}

@article{kurnaz2025joint,
  title={Joint Optimization of Speaker and Spoof Detectors for Spoofing-Robust Automatic Speaker Verification},
  author={Kurnaz, O{\u{g}}uzhan and Mishra, Jagabandhu and Kinnunen, Tomi H and Hanil{\c{c}}i, Cemal},
  journal={arXiv preprint arXiv:2510.01818},
  year={2025}
}

@inproceedings{kurnaz24_asvspoof,
  title     = {Spoofing-robust speaker verification using parallel embedding fusion: {BTU} speech group's approach for {ASV}spoof5 Challenge},
  author    = {Oğuzhan Kurnaz and Selim Can Demirtaş and Aykut Büker and Jagabandhu Mishra and Cemal Hanilçi},
  booktitle = {The Automatic Speaker Verification Spoofing Countermeasures Workshop (ASVspoof 2024)},
  pages     = {138--143},
  doi       = {10.21437/ASVspoof.2024-20},
}

\end{document}